\documentclass[conference]{IEEEtran}
\IEEEoverridecommandlockouts
\usepackage{cite}
\usepackage{amsmath,amssymb,amsfonts}
\usepackage{booktabs}
\usepackage{array,tabularx,multirow}
\usepackage{graphicx}
\usepackage{textcomp}
\usepackage{xcolor}
\usepackage{url}
\usepackage{enumitem}
\usepackage{float}
\usepackage{placeins}
\usepackage{balance}
\usepackage{siunitx}
\usepackage{tikz}
\usetikzlibrary{arrows.meta,positioning,fit,backgrounds,shapes,calc}
\usepackage{amsthm}
\usepackage[linesnumbered,ruled,vlined]{algorithm2e}
\usepackage[hidelinks]{hyperref}

\definecolor{revisionblue}{RGB}{0,0,0}

\newcommand{\stitle}[1]{\par\addvspace{1ex}\noindent\textbf{#1}}
\newcommand{\nop}[1]{}

\newcolumntype{L}[1]{>{\raggedright\arraybackslash}p{#1}}
\newcolumntype{C}[1]{>{\centering\arraybackslash}p{#1}}
\newcolumntype{R}[1]{>{\raggedleft\arraybackslash}p{#1}}

\SetCommentSty{mycommfont}
\SetKwRepeat{Do}{do}{while}
\DeclareMathOperator*{\argmin}{arg\,min}

\newtheorem{lemma}{Lemma}[section]

\theoremstyle{remark}

\theoremstyle{definition}
\newtheorem{definition}{Definition}

\graphicspath{{figures/}}

\title{UniGraph: A Unified Graph-based Index for Dynamic Hybrid Dense-Sparse Vector Retrieval}

\author{
	Keyu Chen{\small$^{1}$}, Xiaoyao Zhong{\small$^{1}$}, Jiabao Jin{\small$^{1}$},   Peng Cheng{\small$^{1}$}, Wangze Ni{\small$^{2}$}, Jingkuan Song{\small$^{1}$}, Heng Tao Shen{\small$^{1}$}\\
	\fontsize{10}{10}\itshape
	$^{1}$Tongji University, Shanghai, China;  $^{2}$Zhejiang University, Hangzhou, China\\
	\fontsize{9}{9}\upshape
	2634115@tongji.edu.cn,  xiaoyao.zhong@hotmail.com,
	jiabaojin0723@gmail.com, cspcheng@tongji.edu.cn,\\
	niwangze@zju.edu.cn, jingkuan.song@gmail.com, shenhengtao@hotmail.com
}

\begin{document}

\maketitle

\begin{abstract}
Hybrid dense-sparse vector retrieval combines semantic signals from dense representations with lexical signals from sparse representations, typically ranking items by a query-dependent weighted sum of dense and sparse vector distances. Existing systems commonly search separate dense and sparse vector indexes and rerank the union of their candidates, incurring two retrieval passes and potentially discarding items that rank highly only after the semantic and lexical signals are combined. Constructing a graph-based index directly under a hybrid distance can mitigate this premature candidate loss. However, because varying the hybrid weight alters the underlying neighborhood relationships, a single static graph built at a fixed weight cannot accommodate dynamic query-specific weights. Therefore, we propose a Unified Hybrid Graph (UniGraph), which merges neighbor sets computed at multiple hybrid weights into a single bounded-degree hybrid graph, enabling hybrid vector retrieval with different query-specific weights without index reconstruction. We further develop an adaptive weight-guided search framework that chooses between sparse- and dense-guided entry points according to the query weight, together with a dense-first hybrid distance pruning strategy that avoids unnecessary exact sparse distance computations. Extensive experiments show that UniGraph provides a favorable Recall-QPS trade-off across different query weights. In the mixed-weight setting, UniGraph delivers 27.6\%$\sim$52.3\% higher QPS than the fastest baseline across datasets at Recall@100 $\geq0.95$. UniGraph has been deployed at scale in Ant Group's production environment, where hybrid retrieval with query-dependent weights is in strong demand.
\end{abstract}


\section{Introduction}
\label{sec:introduction}

\begin{figure*}[t]
\centering
\includegraphics[width=\textwidth,trim={0pt 135pt 0pt 105pt},clip]{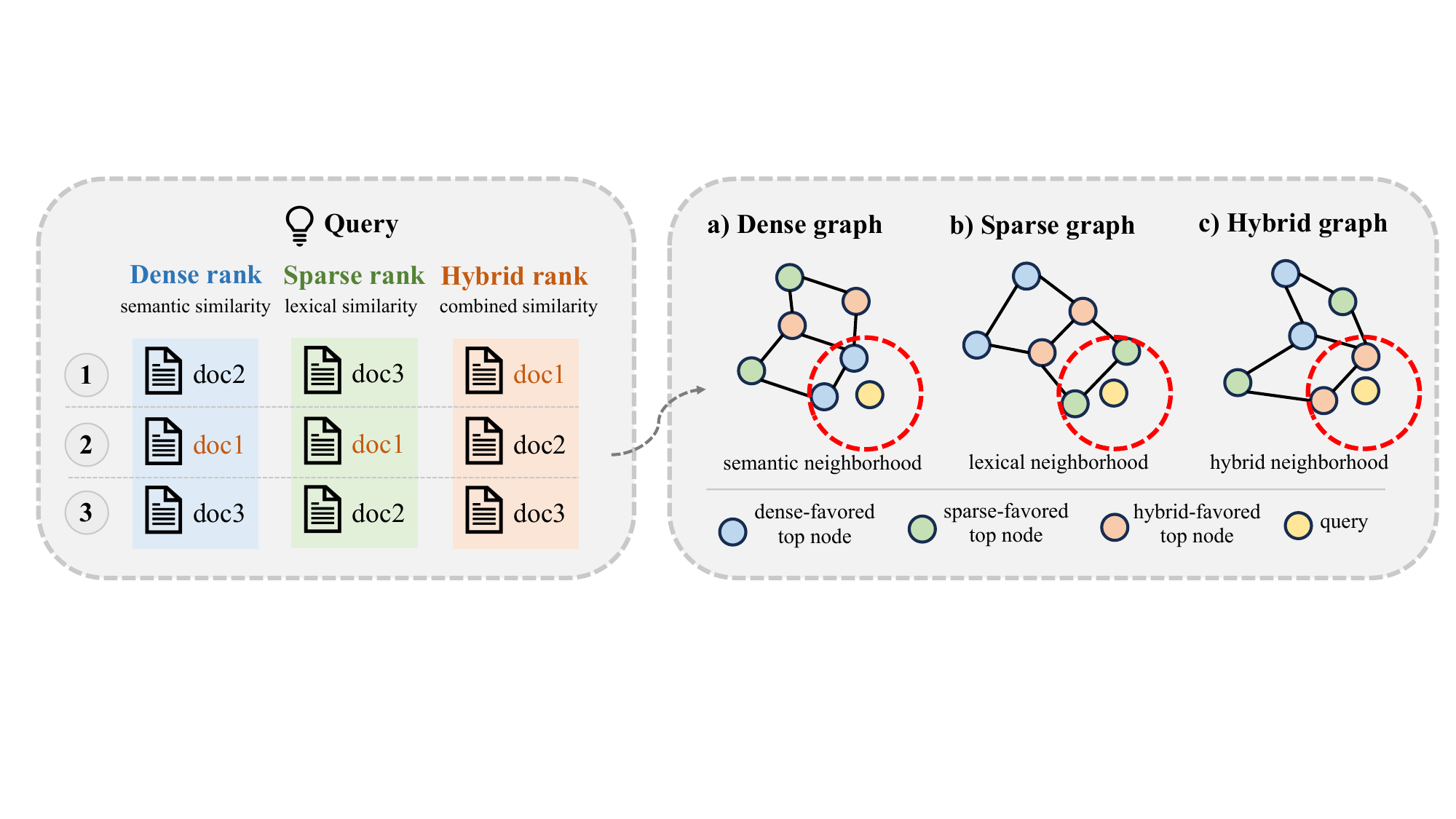}\vspace{-2ex}
\caption{Motivating example for unified dense-sparse hybrid retrieval. The left panel shows that dense, sparse, and hybrid retrieval can produce different top-ranked results for the same query. In this example, doc1 is not ranked first by either dense or sparse retrieval, but becomes the top result under hybrid ranking. The right panel shows three graph neighborhoods built from dense, sparse, and hybrid similarities. Although the three neighborhoods have different highest-ranked nodes, their top-neighbor sets still overlap substantially.}
\label{fig:motivation-example}\vspace{-2ex}
\end{figure*}

The growing volume of unstructured data and the increasing use of retrieval-driven AI applications have established vector search as a fundamental component of modern information systems. Retrieval-augmented generation (RAG), for example, relies on vector search to retrieve relevant information from external document collections and utilizes the retrieved information to provide large language models with additional context~\cite{lewis2020rag,gao2024ragsurvey}. In vector search, data objects and queries are encoded as high-dimensional vectors, and the objects are ranked according to their distances or similarities to the query.

Vector search commonly uses either dense or sparse representations. Dense vectors effectively capture semantic similarity but may not preserve exact lexical information such as rare terms or identifiers~\cite{sciavolino2021simple}. Sparse vectors retain term-level information and are therefore well suited to exact lexical matching~\cite{robertson2009bm25,formal2021splade}. By combining the complementary semantic and lexical signals from dense and sparse representations, hybrid retrieval can achieve higher retrieval effectiveness than dense-only or sparse-only retrieval~\cite{luan2021sparse}. Hybrid dense-sparse vector retrieval has also seen growing adoption in practical information retrieval systems~\cite{sawarkar2024blendedrag}.

In hybrid dense-sparse vector retrieval, each item $x$ in the base dataset is represented by a dense vector $x^d$ and a sparse vector $x^s$. Given a query $q$ with its dense representation $q^d$ and sparse representation $q^s$, the hybrid distance $D_h(q,x;\alpha)$ is defined as a weighted sum of the dense distance $D_d(q^d,x^d)$ and the sparse distance $D_s(q^s,x^s)$:
\begin{equation}
    D_h(q,x;\alpha)=\alpha\cdot D_d(q^d,x^d)+(1-\alpha)\cdot D_s(q^s,x^s),
    \label{eq:hybrid-distance}
\end{equation}
where $\alpha\in[0,1]$ controls the trade-off between semantic and lexical matching. This weight is naturally query-dependent. A conceptual question may rely more on dense similarity, while a query containing rare terms or identifiers may require stronger sparse matching. 
In this work, we consider the setting where $\alpha$ is specified at query time, focusing on efficiently supporting retrieval under the resulting query-specific hybrid distance.

%Hybrid retrieval presents three central challenges. First, a hybrid-retrieval framework should support query-level dynamic weights. Different queries may require different dense--sparse trade-offs; therefore, the index cannot be tied to one fixed $\alpha$ or rebuilt for each query. Second, hybrid retrieval must remain efficient. Evaluating the hybrid distance is expensive, since it requires computing both the dense and sparse distances. Third, hybrid retrieval must maintain robust retrieval effectiveness across different query weights. Different weights correspond to different retrieval objectives, and a hybrid index must consistently achieve high recall across these objectives.

A common existing approach to hybrid retrieval is a two-route pipeline that searches dense and sparse indexes independently, merges their candidate sets, and reranks them under the query-specific hybrid distance~\cite{kuzi2020hybrid,wang2021interpolation,bruch2024fusion,chen2024onesparse}. This design entails an effectiveness--efficiency trade-off: for a requested result size ($k$), retrieving fewer candidates from each index risks missing some of the final hybrid top-$k$ results, whereas retrieving more candidates increases the costs of both searches and reranking. Alternatively, a graph constructed directly under a certain $D_h$ is tailored to the hybrid query weight used during construction and cannot maintain stable retrieval performance under dynamically varying query weights~\cite{zhang2024hybridann}.

Our key observation is that \textit{dense and sparse vectors are complementary representations of the same underlying objects.} Although they emphasize different aspects of relevance, both are derived from the same content and therefore induce correlated similarity structures: objects with high dense similarity also tend to have high sparse similarity. Figure~\ref{fig:motivation-example} illustrates this observation.  A more direct and efficient design is to use a unified index in which dense and sparse signals are combined during graph construction and traversal. 

With this observation, we propose the \textit{Unified Hybrid Graph} (UniGraph), a graph-based index for dense-sparse hybrid retrieval with dynamic query-specific weights. UniGraph constructs a single bounded-degree graph by merging neighborhoods sampled at multiple  $\alpha$ values for various dense-sparse trade-offs. The resulting graph retains edges useful across different dense-sparse trade-offs, allowing the same graph to support different query weights without reconstruction. At query time, our search algorithm first obtains a set of entry points according to the query-specific $\alpha$, aligning graph initialization with the query's dense-sparse preference. It then traverses the unified graph and ranks results using the query-specific hybrid distance. To further improve search efficiency, we develop a dense-first hybrid distance pruning strategy during traversal, which computes the dense component first and utilizes sparse vector norms to determine whether exact sparse distance evaluation can be skipped.

In summary, this paper makes the following contributions:
\begin{itemize}
    \item We propose UniGraph, a unified graph index for dense-sparse hybrid retrieval with dynamic query-specific weights. Its $\alpha$-cover construction merges neighborhoods sampled at multiple dense-sparse trade-off weights into a bounded-degree graph, allowing a single hybrid graph to support different query-specific weights without reconstruction.

    \item We develop an $\alpha$-adaptive search strategy in UniGraph. It routes each query through a sparse-guided or dense-guided execution path according to trade-off value $\alpha$ and applies dense-first hybrid distance pruning to avoid unnecessary exact sparse distance computations.

    \item We conduct extensive experiments on five real-world datasets. At Recall@100 $\geq0.95$, UniGraph delivers 27.6\%$\sim$52.3\% higher QPS than the fastest baseline on each dataset under mixed weights, and $1.18\times$$\sim$$2.85\times$ QPS of the fastest baseline across the fixed weights.
\end{itemize}

The remainder of this paper is organized as follows: \S\ref{sec:preliminaries} defines the hybrid retrieval problem and discusses existing two-route and graph-based solutions. \S\ref{sec:uhg} presents the construction and search procedure of UniGraph, including hybrid distance pruning. \S\ref{sec:guided-uhg-search} presents guided entry-point search in UniGraph and $\alpha$-adaptive execution through sparse-guided and dense-guided paths. \S\ref{sec:experimental} reports the experimental setup and results on five datasets. \S\ref{sec:related} reviews related work.

\section{Preliminaries}
\label{sec:preliminaries}

\subsection{Problem Definition}

\subsubsection{Nearest Neighbor Search}

\begin{definition}[$k$-Nearest Neighbor Search ($k$-NNS)]
Let $\mathcal{X}=\{x_i\}_{i=1}^{N}$ be a finite vector dataset with $|\mathcal{X}|=N$. Given a query vector $q$, a distance function $D(q,x_i)$, and a result size $k$, the objective of $k$-nearest neighbor search ($k$-NNS) is to retrieve a size-$k$ subset of $\mathcal{X}$ whose vectors have the smallest distances to $q$. The exact result is defined as
\begin{equation}
    R_D(q,\mathcal{X},k)=
    \argmin_{\substack{S\subseteq\mathcal{X}\\ |S|=k}}
    \sum_{x_i\in S} D(q,x_i).
    \label{eq:knns}
\end{equation}
\end{definition}

\begin{definition}[$k$-Approximate Nearest Neighbor Search ($k$-ANNS)]
Given a finite vector dataset $\mathcal{X}=\{x_i\}_{i=1}^{N}$, a query vector $q$, a distance function $D(q,x_i)$, and a result size $k$, the objective of $k$-approximate nearest neighbor search ($k$-ANNS) is to efficiently retrieve a size-$k$ approximate result set $\widehat{R}_D(q,\mathcal{X},k)$ that closely matches the exact $k$-NNS result $R_D(q,\mathcal{X},k)$ without exhaustively evaluating $D(q,x_i)$ over all $x_i\in\mathcal{X}$. The retrieval quality of $\widehat{R}_D(q,\mathcal{X},k)$ is measured by its recall against the exact result:
\begin{equation}
    \mathrm{Recall@}k(q;D)=
    \frac{\left|\widehat{R}_D(q,\mathcal{X},k)\cap R_D(q,\mathcal{X},k)\right|}{k}.
    \label{eq:recall}
\end{equation}
\end{definition}

\subsubsection{Dense and Sparse Retrieval}

In the dense-sparse retrieval setting, each base item $x_i\in\mathcal{X}$ is represented as a dense-sparse vector pair $x_i=(x_i^d,x_i^s)$. A query is written as $q=(q^d,q^s)$, where $q^d$ and $q^s$ represent the dense and sparse vectors of query $q$, respectively.

A dense base vector $x^d$ is obtained by encoding an item in the base dataset using a dense encoding model, such as DPR~\cite{karpukhin2020dpr}, Sentence-BERT~\cite{reimers2019sentencebert}, or BGE-M3~\cite{chen2024bgem3}. Dense retrieval ranks dense base vectors using distance measures such as inner-product distance, cosine distance, or Euclidean distance. Let $D_d(q^d,x_i^d)$ denote the dense distance between query vector $q^d$ and base vector $x_i^d$. The exact dense top-$k$ result $R_d(q,\mathcal{X},k)$ is the size-$k$ subset $S\subseteq\mathcal{X}$ that minimizes the total dense distance:
\begin{equation}
    R_d(q,\mathcal{X},k)=
    \argmin_{\substack{S\subseteq\mathcal{X}\\ |S|=k}}
    \sum_{x_i\in S} D_d(q^d,x_i^d).
    \label{eq:dense-retrieval}
\end{equation}

A sparse base vector $x^s$ has high dimensionality, with most dimensions equal to zero. Nonzero entries correspond to vocabulary terms and preserve lexical information. Sparse base vectors can be produced by term-weighting models such as BM25~\cite{robertson2009bm25} or learned sparse encoders such as SPLADE~\cite{formal2021splade,formal2022sparsehardnegative,lassance2022spladeefficiency,formal2024sparseir}. Let $T(q^s)$ and $T(x_i^s)$ denote the sets of nonzero terms in the sparse query and base vector, respectively, and let $q_t^s$ and $x_{i,t}^s$ be their weights for term $t$. The sparse inner-product score $s_s(q^s,x_i^s)$ between query vector $q^s$ and base vector $x_i^s$ is defined as
\begin{equation}
    s_s(q^s,x_i^s)=\sum_{t\in T(q^s)\cap T(x_i^s)}
    q^s_t x^s_{i,t}.
    \label{eq:sparse-score}
\end{equation}
The exact sparse top-$k$ result $R_s(q,\mathcal{X},k)$ is the size-$k$ subset $S\subseteq\mathcal{X}$ that maximizes the total sparse score:
\begin{equation}
    R_s(q,\mathcal{X},k)=
    \operatorname*{arg\,max}_{\substack{S\subseteq\mathcal{X}\\ |S|=k}}
    \sum_{x_i\in S} s_s(q^s,x_i^s).
    \label{eq:sparse-retrieval}
\end{equation}

\subsubsection{Hybrid Dense-Sparse Retrieval}

We use the hybrid distance defined in Eq.~\eqref{eq:hybrid-distance}, with the query-level weight $\alpha\in[0,1]$ controlling the dense-sparse trade-off.
Sparse retrieval uses inner-product scores, which we convert to distances as $D_s(q^s,x_i^s)=1-s_s(q^s,x_i^s)=1-\langle q^s,x_i^s\rangle$. To use a consistent distance formulation for the dense and sparse components, we also adopt inner-product distance for dense retrieval, i.e., $D_d(q^d,x_i^d)=1-\langle q^d,x_i^d\rangle$.
In this definition, a smaller $\alpha$ value places greater weight on the sparse distance, whereas a larger $\alpha$ value places greater weight on the dense distance.

\begin{definition}[Hybrid Dense-Sparse Retrieval]
Given a dataset $\mathcal{X}$ of base vector pairs, a query vector pair $q$, a trade-off weight $\alpha$, and a result size $k$, the hybrid dense-sparse retrieval problem is to find the top-$k$ base vector pairs in $\mathcal{X}$ with the smallest hybrid distances:
\begin{equation}
    R_h(q,\mathcal{X},k;\alpha)=
    \argmin_{\substack{S\subseteq\mathcal{X}\\ |S|=k}}
    \sum_{x_i\in S} D_h(q,x_i;\alpha).
    \label{eq:hybrid-retrieval}
\end{equation}
\end{definition}

At scale, we address this objective as a $k$-ANNS problem under $D_h$. Let $\widehat{R}_h(q,\mathcal{X},k;\alpha)$ denote the returned top-$k$ result set; its recall is computed using Eq.~\eqref{eq:recall} against the exact result $R_h(q,\mathcal{X},k;\alpha)$. The challenge studied in this paper is to compute high-recall approximations $\widehat{R}_h(q,\mathcal{X},k;\alpha)$ efficiently for query-level $\alpha$ values without rebuilding the index for each query. Table~\ref{tab:notation} summarizes the main symbols used in the paper.

\begin{table}[t]
\centering
\caption{Summary of notation.}
\label{tab:notation}
\small
\renewcommand{\arraystretch}{1.08}
\setlength{\tabcolsep}{3pt}
\begin{tabular}{@{}p{0.29\linewidth}p{0.68\linewidth}@{}}
\toprule
Symbol & Description \\
\midrule
$\mathcal{X}$, $N$ & Base dataset and its size \\
$x_i=(x_i^d,x_i^s)$ & Dense-sparse base vector pair \\
$q=(q^d,q^s)$ & Dense-sparse query vector pair \\
$D_d$, $D_s$, $D_h$ & Dense, sparse, and hybrid distances \\
$\alpha$ & Hybrid weight \\
$R_d$, $R_s$, $R_h$ & Dense, sparse, and hybrid top-$k$ results \\
$\mathcal{A}$ & Hybrid weights sampled for graph construction \\
$\delta$ & Sampling step for hybrid weights \\
$k_g$ & Number of neighbors retained per sampled weight \\
$\mathcal{N}_i^\alpha$ & Top-$k_g$ neighbors of $x_i$ under weight $\alpha$ \\
$\mathcal{U}_i$ & Union of neighbor sets of $x_i$ over sampled weights \\
$f_i(x_j)$ & Selection frequency of $x_j$ as a neighbor of $x_i$ \\
$M$ & Maximum graph out-degree \\
$N_G(x)$ & Outgoing neighbor set of $x$ in UniGraph \\
$\mathcal{C}$, $\mathcal{R}$ & Candidate and result sets in graph search \\
$k$, $\mathcal{R}_k$ & Number of returned results and final top-$k$ result set \\
$\mathcal{V}_{\mathrm{seen}}$ & Vertices seen during graph search \\
$ef_c$, $ef_s$ & Construction and search beam sizes \\
$\gamma$ & Hybrid-distance pruning scale \\
\bottomrule
\end{tabular}
\end{table}

\begin{figure*}[!t]\vspace{-2ex}
\centering
\includegraphics[width=\textwidth,trim={20pt 5pt 2pt 5pt},clip]{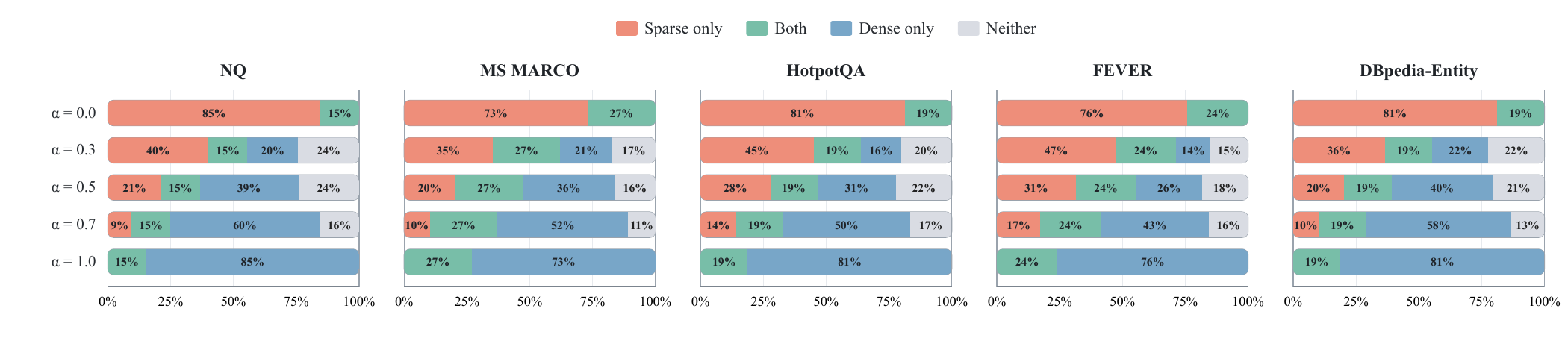}\vspace{-3ex}
\caption{Source composition of the exact hybrid result set $R_h$ for $k=100$. Each bar partitions the items in $R_h$ according to their membership in $R_s$ and $R_d$: only in $R_s$, only in $R_d$, in both sets, or in neither set. The \emph{Both} segment remains relatively stable across $\alpha$ values because the items it contains rank highly in both dense and sparse retrieval and therefore also tend to rank highly under the hybrid distance.}
\label{fig:hybrid-composition}\vspace{-2ex}
\end{figure*}

\subsection{Existing Solutions}

\textit{Two-route retrieval.} A common way to support hybrid retrieval is to run dense and sparse retrieval as two independent routes~\cite{kuzi2020hybrid,wang2021interpolation,bruch2024fusion,chen2024onesparse}. For a query $q$, a dense vector index retrieves top candidates ranked by the dense distance $D_d$, while a sparse vector index retrieves top candidates ranked by the sparse distance $D_s$. The union is then reranked by computing $D_h(q,x;\alpha)$ for each candidate $x$ to produce the final top-$k$~\cite{wang2021interpolation,bruch2024fusion}. The two-route design imposes no restrictions on the choice of dense or sparse indexes. The number of candidates retrieved from each index can be adjusted to balance recall and retrieval efficiency. Retrieving more candidates can improve recall but increases search and reranking costs, whereas retrieving fewer candidates reduces these costs but risks missing hybrid top-$k$ results~\cite{chen2024onesparse}.

\textit{Fixed-weight hybrid graph.} Another approach is to construct a proximity graph using the hybrid distance with a fixed weight $\alpha$, such as $\alpha=0.5$~\cite{zhang2024hybridann}. Because nearest-neighbor relationships can change with the hybrid weight, a graph constructed at a fixed $\alpha$ may lack connections between items that are close under other query weights. When the query weight differs from the weight used for graph construction, the search may need to explore more vertices to reach the nearest neighbors. This can increase query cost or reduce recall for a fixed search budget. Although this approach can be improved by first performing coarse retrieval with a dense or sparse index and then conducting fine-grained search on the hybrid graph~\cite{zhang2024hybridann}, it still fundamentally relies on a graph constructed with a fixed hybrid weight and therefore cannot adapt to query-dependent hybrid weights.

\section{Unified Hybrid Graph}
\label{sec:uhg}

\subsection{Motivation}
\label{sec:motivation}

Dense and sparse vectors are paired representations of the same items in the base dataset. They are partially correlated: not identical, but not independent. This partial correlation creates both the problem and the opportunity. The problem is that two-route retrieval may miss hybrid top-$k$ results when an item does not rank highly in either vector modality. The opportunity is that neighborhoods under nearby $\alpha$ values overlap, allowing a unified graph to reuse edges across multiple hybrid weights.

Figure~\ref{fig:hybrid-composition} analyzes the source composition of the exact hybrid top-$k$ result sets across five real-world datasets. For $k=100$, we classify each item in the exact hybrid result set according to whether it also appears in the exact dense and sparse result sets. At intermediate weights, both dense-only and sparse-only results account for substantial portions of the hybrid top-100, indicating that neither component overwhelmingly dominates the hybrid ranking. This suggests that, for the datasets shown, the dense and sparse distance scales are sufficiently compatible. The \emph{Neither} portion shows that some hybrid top-$k$ results are absent from both single-modality result sets. On dataset NQ at $\alpha=0.3$ and $0.5$, this portion reaches 24\%. These omissions limit the recall of two-route retrieval.

Figure~\ref{fig:hybrid-composition} also suggests why a unified graph is feasible. As $\alpha$ increases from 0 to 1, the contribution of the two modalities changes accordingly, while the portion shared by both rankings remains stable. In other words, hybrid neighborhoods are weight-dependent but still related: different weights favor different neighbors, yet part of the neighborhood structure can be reused. A graph constructed at a fixed $\alpha$ value captures only the neighbor relations under that weight. This motivates the $\alpha$-cover construction of UniGraph, which combines neighbor relations from multiple $\alpha$ values, retaining weight-specific connections while sharing their common structure in a single graph.

\stitle{Why Is Discrete $\alpha$ Sampling Effective?}
The effectiveness of discrete sampling depends on how the hybrid neighborhood changes with $\alpha$.
For a fixed center base vector pair $x_i$ and a candidate neighbor $x_j$, the hybrid distance can be rewritten as
{\small\begin{equation}
    D_h(x_i,x_j;\alpha)
    =D_s(x_i^s,x_j^s)
    +\alpha\big(D_d(x_i^d,x_j^d)-D_s(x_i^s,x_j^s)\big).
    \label{eq:hybrid-distance-alpha}
\end{equation}}
This equation shows that the hybrid distance between two base vector pairs changes linearly with $\alpha$. For a fixed center base vector pair $x_i$, each of the remaining $N-1$ base vector pairs defines a linear hybrid distance function of $\alpha$. When two candidates are compared, their relative order can change only at a crossing of their distance functions. Only crossings within $\alpha\in[0,1]$ can affect the candidate ranking, and the partial correlation between dense and sparse distances further limits the number of such crossings. Many candidate pairs have the same relative order under both the dense and sparse distances. For such pairs, this order is preserved under any convex combination of the two distances. Even among the remaining crossings, those within the selected neighbors or within the unselected candidates change only their internal order. For a per-weight neighbor-set size ($k_g$), the neighbor set changes only when a candidate crosses the top-$k_g$ boundary. Only a relatively small number of pairwise crossings can change the top-$k_g$ neighbor set. Between consecutive change points, the neighbor set remains unchanged. Therefore, discrete $\alpha$ sampling can effectively capture neighborhood structures across the continuous weight range. Accordingly, to construct UniGraph, we use a fixed sampling step $\delta$ to select a finite set of $\alpha$ values instead of finding all change points explicitly.

\subsection{UniGraph Construction}
\label{sec:uhg-construction}

\begin{figure*}[!t]\vspace{-1ex}
\centering
\includegraphics[width=\textwidth,trim={0pt 0pt 0pt 0pt},clip]{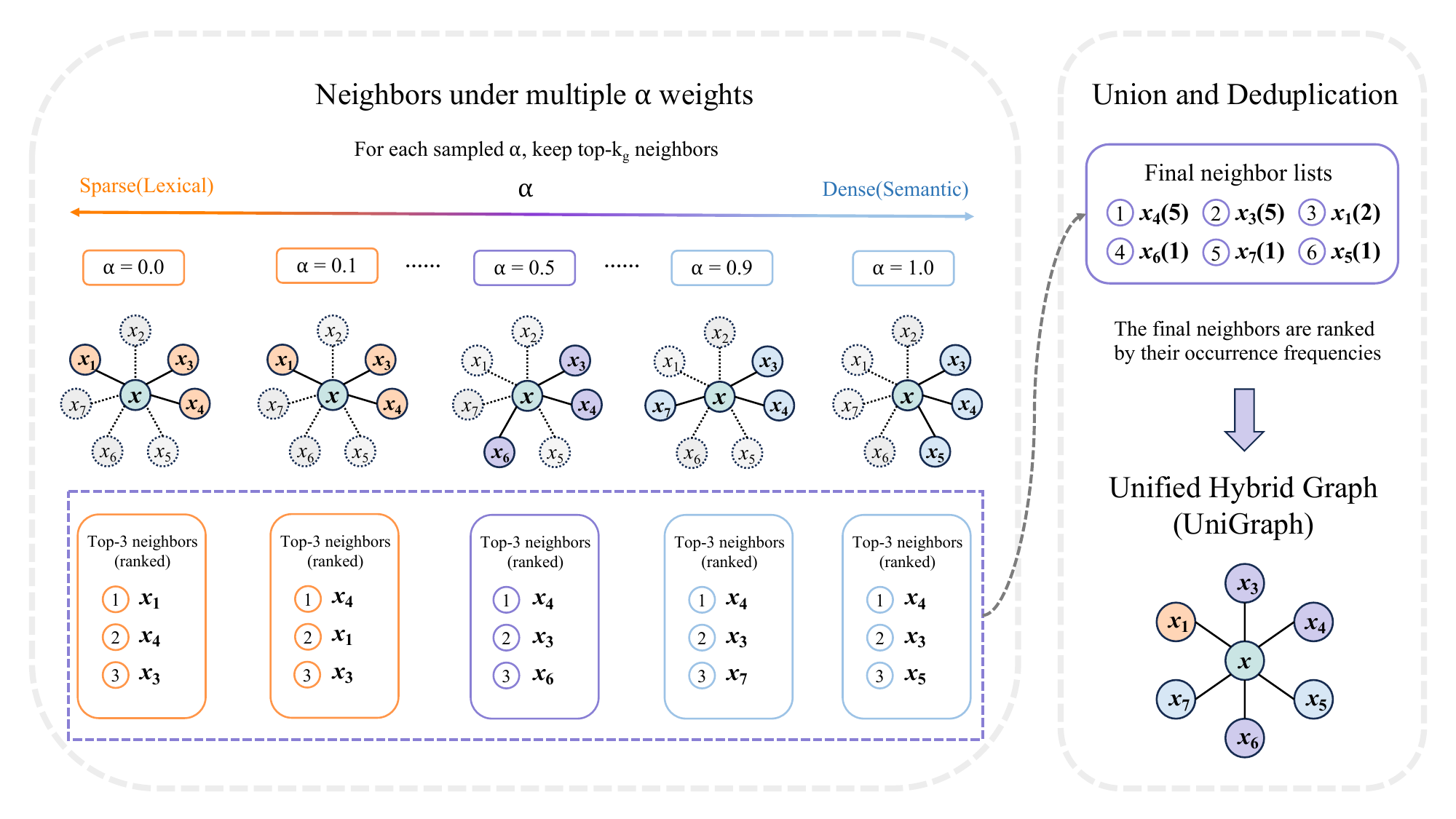}\vspace{-2ex}
\caption{Overview of UniGraph construction for one center base vector pair $x$. Across sampled $\alpha$ values ranging from the sparse (lexical) endpoint to the dense (semantic) endpoint, the construction retains the top-$k_g$ hybrid neighbors of $x$ at each weight; the illustration uses $k_g=3$. The per-$\alpha$ lists are merged and deduplicated, and the number in parentheses indicates each candidate's occurrence frequency. Ranking the candidates by this frequency produces a single UniGraph neighborhood that favors neighbors selected across multiple dense-sparse trade-offs.}
\label{fig:uhg-construction-overview}
\end{figure*}

Given a sampling step $\delta$, we discretize $[0,1]$ into $\mathcal{A}=\{0,\delta,2\delta,\ldots,1\}$ and compute the top-$k_g$ hybrid neighbors of $x_i$ for each $\alpha\in\mathcal{A}$. We denote this per-$\alpha$ neighbor set by
\begin{equation}
    \mathcal{N}_i^\alpha=
    \argmin_{\substack{S\subseteq\mathcal{X}\setminus\{x_i\}\\ |S|=k_g}}
    \sum_{x_j\in S}D_h(x_i,x_j;\alpha).
    \label{eq:per-alpha-neighbors}
\end{equation}
The $\alpha$-cover neighbor set is
\begin{equation}
    \mathcal{U}_i=\bigcup_{\alpha\in\mathcal{A}}
    \mathcal{N}_i^\alpha.
    \label{eq:alpha-cover}
\end{equation}
This set approximates the union of the top-$k_g$ neighbor sets induced by all hybrid weights $\alpha\in[0,1]$. In our experiments, we find that a sampling step of $\delta=0.1$ is sufficient to capture neighborhood changes, and using a smaller step yields little additional improvement.

After collecting the per-$\alpha$ top-$k_g$ sets, we perform an $\alpha$-frequency merge. For each candidate $x_j$, let
\begin{equation}
    f_i(x_j)=
    \sum_{\alpha\in\mathcal{A}}
    \mathbf{1}\!\left[
    x_j\in \mathcal{N}_i^\alpha
    \right],
    \label{eq:alpha-frequency}
\end{equation}
where $\mathbf{1}[\cdot]$ is the indicator function, equal to 1 when its condition holds and 0 otherwise.
We represent UniGraph as a directed graph $G=(V,E)$, where $V=\mathcal{X}$ is the vertex set and $E$ is the edge set. The final outgoing neighbor set of $x_i$ in $G$ retains the candidates with the highest frequencies:
\begin{equation}
    N_G(x_i)=
    \operatorname*{arg\,max}_{\substack{S\subseteq\mathcal{U}_i\\
    |S|=\min\{M,|\mathcal{U}_i|\}}}
    \sum_{x_j\in S}f_i(x_j).
    \label{eq:alpha-frequency-merge}
\end{equation}
This caps the outgoing degree at $M$ by favoring neighbors that are repeatedly selected across sampled hybrid weights rather than discarding candidates at random. A higher occurrence frequency indicates that the neighbor remains important under more dense-sparse trade-offs. Figure~\ref{fig:uhg-construction-overview} illustrates the $\alpha$-cover construction workflow. Overall, the construction samples multiple hybrid weights, computes the top-$k_g$ neighbors at each weight, merges the resulting neighbor sets, and ranks the candidates by occurrence frequency to obtain the final neighbor list used by UniGraph.

\begin{algorithm}[tbp]
\small
\DontPrintSemicolon
\caption{UniGraph Construction}
\label{alg:uhg-construction}
\KwIn{base dataset $\mathcal{X}$, sampled $\alpha$ values $\mathcal{A}$, number of neighbors kept for each $\alpha$ $k_g$, maximum out-degree $M$.}
\KwOut{$G=(V,E)$.}
$V\leftarrow\mathcal{X}$; $E\leftarrow\emptyset$\;
\ForEach{$x_i\in\mathcal{X}$}{
    $f_i(\cdot)\leftarrow 0$; $\mathcal{U}_i\leftarrow\emptyset$\;
    \ForEach{$\alpha\in\mathcal{A}$}{
        $\displaystyle
        \mathcal{N}_i^\alpha\leftarrow
        \argmin_{\substack{S\subseteq\mathcal{X}\setminus\{x_i\}\\ |S|=k_g}}
        \sum_{x_j\in S}D_h(x_i,x_j;\alpha)$\;
        $\mathcal{U}_i\leftarrow\mathcal{U}_i\cup\mathcal{N}_i^\alpha$\;
        \ForEach{$x_j\in\mathcal{N}_i^\alpha$}{
            $f_i(x_j)\leftarrow f_i(x_j)+1$\;
        }
    }
    $\displaystyle
    N_G(x_i)\leftarrow
    \operatorname*{arg\,max}_{\substack{S\subseteq\mathcal{U}_i\\
    |S|=\min\{M,|\mathcal{U}_i|\}}}
    \sum_{x_j\in S}f_i(x_j)$\;
    $E\leftarrow E\cup\{(x_i,x_j)\mid x_j\in N_G(x_i)\}$\;
}
$G\leftarrow(V,E)$\;
\KwRet{$G$}\;
\end{algorithm}

Algorithm~\ref{alg:uhg-construction} formalizes the complete $\alpha$-cover construction of UniGraph. It first initializes the vertex and edge sets (Line~1). For each center $x_i$ and each sampled $\alpha$, the algorithm selects the $k_g$ other base vector pairs with the smallest hybrid distances to $x_i$, inserts them into $\mathcal{U}_i$, and increments $f_i(x_j)$ once for every selected $x_j$ (Lines~2--8). Consequently, after all sampled weights have been processed, $\mathcal{U}_i$ and $f_i$ are exactly the union and occurrence counts in Eqs.~\eqref{eq:alpha-cover} and~\eqref{eq:alpha-frequency}. The algorithm then keeps the $\min\{M,|\mathcal{U}_i|\}$ candidates with the largest frequencies as $N_G(x_i)$ and adds a directed edge $(x_i,x_j)$ for each retained neighbor (Lines~9--10). Repeating these operations for every $x_i$ completes $E$, after which the algorithm returns $G=(V,E)$ (Lines~11--12).

\begin{algorithm}[!t]
\small
\DontPrintSemicolon
\caption{UniGraph Search}
\label{alg:uhg-search}
\KwIn{$q$, $G=(V,E)$, entry-point set $E_q$, top-$k$, $ef_s$, hybrid weight $\alpha$, pruning scale $\gamma\in[0,1]$.}
\KwOut{$\mathcal{R}_k$.}
$\mathcal{C}\leftarrow\mathcal{R}\leftarrow
\{(e,D_h(q,e;\alpha))\mid e\in E_q\}$;
$\mathcal{V}_{\mathrm{seen}}\leftarrow E_q$\;
$n_q\leftarrow\|q^s\|_2$\;
$\tau\leftarrow\max_{(v,D_v)\in\mathcal{R}}D_v$\;
\While{$\mathcal{C}\neq\emptyset$}{
    $(x,D_x)\leftarrow
    \argmin_{(v,D_v)\in\mathcal{C}}D_v$;
    $\mathcal{C}\leftarrow\mathcal{C}\setminus\{(x,D_x)\}$\;
    \If{$|\mathcal{R}|=ef_s$ and $D_x>\tau$}{
        break\;
    }
    \ForEach{$y\in N_G(x)$}{
        \If{$y\in\mathcal{V}_{\mathrm{seen}}$}{
            continue\;
        }
        $\mathcal{V}_{\mathrm{seen}}\leftarrow\mathcal{V}_{\mathrm{seen}}\cup\{y\}$\;
        $D_{d,y}\leftarrow D_d(q^d,y^d)$\;
        \If{$|\mathcal{R}|=ef_s$}{
            $\widehat{D}_y\leftarrow\alpha D_{d,y}+(1-\alpha)(1-\gamma n_q\|y^s\|_2)$\;
            \If{$\widehat{D}_y\geq\tau$}{
                continue\;
            }
        }
        $D_{s,y}\leftarrow D_s(q^s,y^s)$\;
        $D_y\leftarrow\alpha D_{d,y}+(1-\alpha)D_{s,y}$\;
        \If{$|\mathcal{R}|<ef_s$ or $D_y<\tau$}{
            $\mathcal{C}\leftarrow\mathcal{C}\cup\{(y,D_y)\}$;
            $\mathcal{R}\leftarrow\mathcal{R}\cup\{(y,D_y)\}$\;
            \If{$|\mathcal{R}|>ef_s$}{
                $(z,D_z)\leftarrow
                \operatorname*{arg\,max}_{(v,D_v)\in\mathcal{R}}D_v$;
                $\mathcal{R}\leftarrow\mathcal{R}\setminus\{(z,D_z)\}$\;
            }
            $\tau\leftarrow\max_{(v,D_v)\in\mathcal{R}}D_v$\;
        }
    }
}
$\displaystyle
\mathcal{R}_k\leftarrow
\argmin_{\substack{S\subseteq\mathcal{R}\\ |S|=k}}
\sum_{(v,D_v)\in S}D_v$\;
\KwRet{$\mathcal{R}_k$}\;
\end{algorithm}

In practice, we use dense and sparse indexes as offline candidate generators during UniGraph construction to avoid all-pairs distance computations. For each base vector pair, we independently retrieve candidate neighbors from the dense and sparse indexes, merge their results into a compact candidate pool, and compute the exact dense and sparse distances only for these candidates. The subsequent construction procedure does not depend on the specific dense and sparse indexes used for candidate generation. The dense and sparse indexes are used only to accelerate UniGraph construction. The per-$\alpha$ neighbor selection and frequency merge in Algorithm~\ref{alg:uhg-construction} are then performed over this pool rather than the full dataset; the remaining construction procedure is unchanged.

\subsection{UniGraph Search}
\label{sec:uhg-search}

Given a query $q$, a hybrid weight $\alpha$, and an entry-point set $E_q$, we perform greedy beam search over UniGraph. Let $ef_s$ denote the search beam size. Unguided UniGraph search starts from a single randomly selected entry point ($E_q=\{e_0\}$), whereas guided UniGraph search uses multiple query-dependent entry points retrieved from a guide index, as detailed in \S\ref{sec:guided-entry-search}. The traversal maintains a candidate set $\mathcal{C}$ ordered by the query-specific hybrid distance and an $ef_s$-sized result set $\mathcal{R}$ containing the best vertices found so far. At each step, the search selects the closest unexpanded vertex from $\mathcal{C}$, visits its graph neighbors, evaluates promising unseen vertices, and updates $\mathcal{C}$ and $\mathcal{R}$. Once $|\mathcal{R}|=ef_s$, it uses the standard beam-search termination condition: traversal stops when the closest remaining candidate is farther than the worst vertex in $\mathcal{R}$. During traversal, UniGraph directly uses the query-specific hybrid distance to order candidates, while supporting any query-level hybrid weight $\alpha\in[0,1]$ without requiring graph reconstruction.

Hybrid distance evaluation can dominate traversal cost because every visited candidate may require computing both dense and sparse distances. Computing the sparse inner-product distance is especially costly because it requires matching query terms with irregularly stored nonzero entries. UniGraph search therefore evaluates the dense component first and combines it with sparse vector norms to obtain an estimated hybrid distance. Based on this estimate, it determines whether exact sparse distance computation is necessary. Base vector norms are query-independent and can be precomputed during index construction and stored with the index; the sparse query norm is computed once and reused for all candidates during the query. Once the dense distance is available, computing the pruning estimate requires only a few scalar operations and no sparse vector matching. We next derive a norm-based lower bound, use it to construct a tunable hybrid-distance estimate, and characterize the resulting pruning guarantee and efficiency--recall tradeoff.

\stitle{Norm-Based Lower Bound}
Let $\|\cdot\|_2$ denote the Euclidean norm. Applying the Cauchy--Schwarz inequality to the sparse inner-product distance defined in \S~\ref{sec:preliminaries}, and noting that $1-\alpha\geq0$, the norm-based sparse estimate and its induced hybrid estimate satisfy
\begin{align}
    L_s(q^s,x^s)
    &=1-\|q^s\|_2\|x^s\|_2 \notag\\
    &\leq 1-\langle q^s,x^s\rangle
    =D_s(q^s,x^s),
    \label{eq:sparse-lower-bound}\\
    L_h(q,x;\alpha)
    &=\alpha D_d(q^d,x^d)
      +(1-\alpha)L_s(q^s,x^s) \notag\\
    &\leq D_h(q,x;\alpha).
    \label{eq:hybrid-lower-bound}
\end{align}
Thus, after computing the dense distance, UniGraph obtains a valid hybrid-distance lower bound using only the vector norms, without matching the nonzero entries of $q^s$ and $x^s$.

\stitle{Scaled Estimate and Pruning Rule}
The lower bound in Eq.~\eqref{eq:hybrid-lower-bound} is safe but can be loose. To control pruning strength, we introduce a pruning scale $\gamma\in[0,1]$ and define the scaled hybrid distance estimate $\widehat{D}_h$ as
{\small\begin{equation}
    \widehat{D}_h(q,x;\alpha,\gamma)
    =\alpha D_d(q^d,x^d)
     +(1-\alpha)\big(1-\gamma\|q^s\|_2\|x^s\|_2\big).
    \label{eq:pruning-bound}
\end{equation}}
Equivalently, $\widehat{D}_h=L_h+(1-\alpha)(1-\gamma)\|q^s\|_2\|x^s\|_2$, so $\gamma$ controls a nonnegative margin above the Cauchy-Schwarz lower bound.

Once $|\mathcal{R}|=ef_s$, let $\tau$ be the current worst hybrid distance in $\mathcal{R}$. UniGraph compares $\widehat{D}_h$ with $\tau$. If $\widehat{D}_h\geq\tau$, it skips the exact sparse distance for $x$; otherwise, it computes $D_s(q^s,x^s)$ and forms the exact hybrid distance.

\begin{figure}[t]
\centering
\includegraphics[width=\columnwidth,trim={275pt 30pt 200pt 30pt},clip]{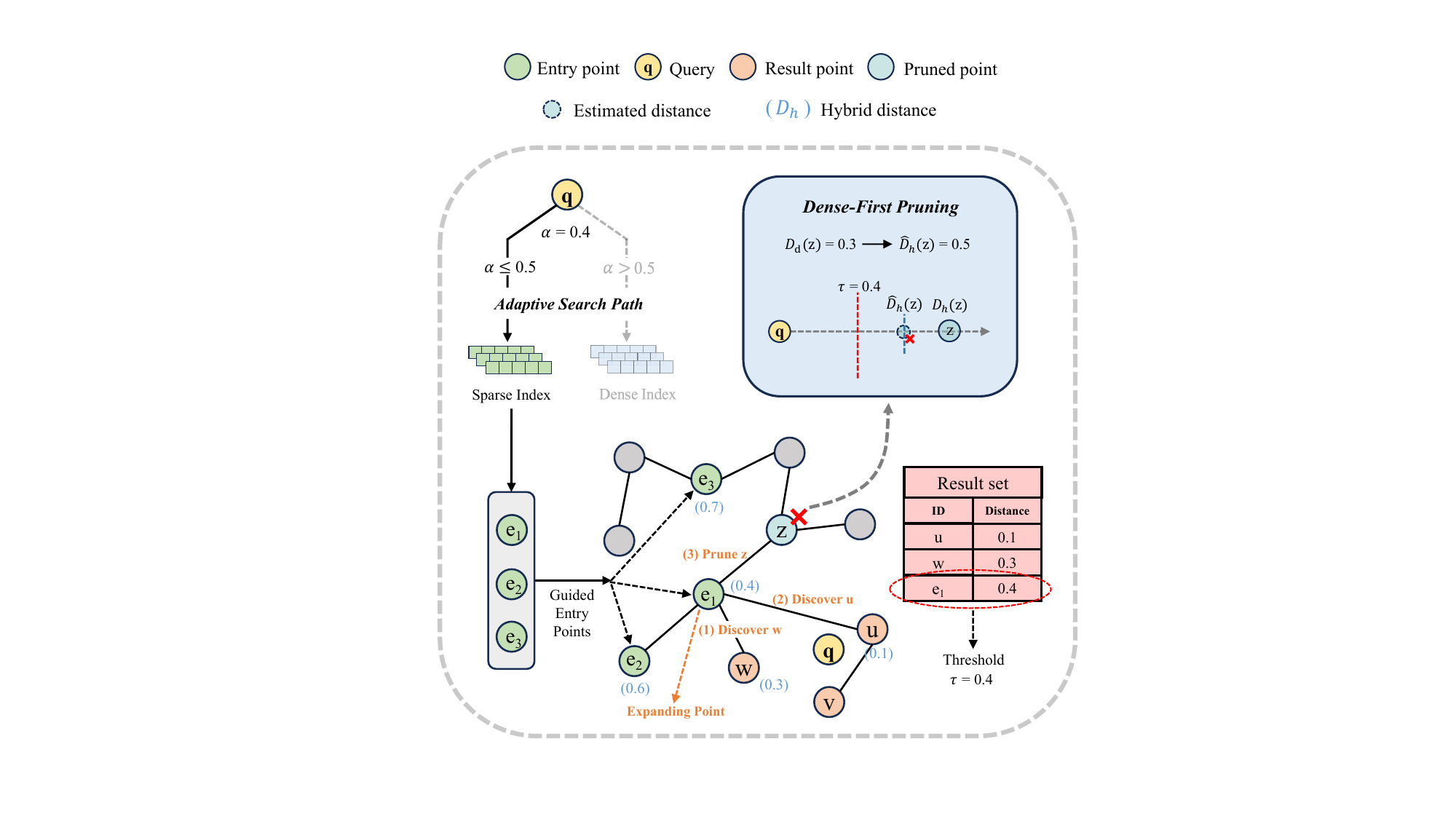}\vspace{-1ex}
\caption{A concrete example of UniGraph search for a query with $\alpha=0.4$. Because $\alpha\leq0.5$, the $\alpha$-adaptive policy selects the sparse guide index, which returns $e_1$, $e_2$, and $e_3$ as entry points. During UniGraph traversal, expanding $e_1$ discovers $w$ and $u$ and updates the result set, whose current worst distance is $\tau=0.4$. For brevity, we omit the fixed query and parameter arguments from the distance notation in this example. For candidate $z$, dense-first pruning uses $D_d(z)=0.3$ to obtain $\widehat{D}_h(z)=0.5$. Since $\widehat{D}_h(z)\geq\tau$, $z$ is pruned without exact sparse distance evaluation. Blue values in parentheses denote exact hybrid distances.}
\label{fig:uhg-search-example}
\end{figure}

\stitle{Pruning Guarantee and Tradeoff}
\begin{lemma}[Safe Dense-First Pruning]
\label{lem:safe-pruning}
Suppose $|\mathcal{R}|=ef_s$ and $\gamma=1$. If $\widehat{D}_h(q,x;\alpha,1)\geq\tau$, then $D_h(q,x;\alpha)\geq\tau$. Hence, the exact-distance beam-search insertion rule would reject $x$, and its exact sparse distance need not be evaluated.
\end{lemma}
\begin{IEEEproof}
When $\gamma=1$, Eq.~\eqref{eq:pruning-bound} gives $\widehat{D}_h=L_h$. By Eq.~\eqref{eq:hybrid-lower-bound}, $L_h\leq D_h$. Hence, $\widehat{D}_h\geq\tau$ implies $D_h\geq\tau$.
\end{IEEEproof}
For $\gamma<1$, the nonnegative margin makes $\widehat{D}_h$ larger and may prune more candidates, but the estimate is no longer guaranteed to lower-bound $D_h$. Smaller $\gamma$ values therefore reduce exact sparse distance evaluations at the risk of changing the search path and affecting recall, whereas larger values are more conservative. We use $\gamma=0.3$ in the main experiments and isolate its throughput effect in the pruning ablation. Figure~\ref{fig:uhg-search-example} illustrates how $\widehat{D}_h$ is compared with the current result-set threshold $\tau$.

Algorithm~\ref{alg:uhg-search} formalizes the complete UniGraph search procedure. It first computes the exact hybrid distance $D_h(q,e;\alpha)$ for each entry $e\in E_q$, inserts the resulting pairs into both $\mathcal{C}$ and $\mathcal{R}$, and marks all entries as seen (Line~1). It computes the query sparse norm $n_q=\|q^s\|_2$ once and initializes $\tau$ to the largest entry distance in $\mathcal{R}$ (Lines~2--3). At each iteration, the algorithm removes from $\mathcal{C}$ the unexpanded pair $(x,D_x)$ with the smallest distance (Lines~4--5); if $\mathcal{R}$ is full and $D_x>\tau$, traversal terminates because no remaining candidate can improve the current result set (Lines~6--7). Otherwise, it scans every outgoing neighbor $y\in N_G(x)$, skips previously seen neighbors, and marks each unseen neighbor immediately so that it is evaluated at most once (Lines~8--11). It first computes the dense component $D_{d,y}$ (Line~12) and, when $\mathcal{R}$ is full, uses $D_{d,y}$, $n_q$, and the stored norm $\|y^s\|_2$ to obtain $\widehat{D}_y$ from Eq.~\eqref{eq:pruning-bound}; if $\widehat{D}_y\geq\tau$, it skips $y$ without computing the exact sparse component (Lines~13--16). If $\mathcal{R}$ is not yet full or $\widehat{D}_y<\tau$, the algorithm computes $D_{s,y}$ and the exact hybrid distance $D_y=\alpha D_{d,y}+(1-\alpha)D_{s,y}$ (Lines~17--18). The result set $\mathcal{R}$ retains at most the $ef_s$ closest vertices found so far, and $\tau$ is the distance of its farthest member (Lines~19--23). If $\mathcal{R}$ is not full, $(y,D_y)$ is pushed directly into both $\mathcal{C}$ and $\mathcal{R}$; once $\mathcal{R}$ is full, the pair is pushed only when $D_y<\tau$, meaning that $y$ is closer than the current farthest member (Lines~19--20). If this insertion makes $|\mathcal{R}|>ef_s$, the algorithm removes the farthest pair and updates $\tau$ to the largest remaining distance (Lines~21--23). When traversal terminates or $\mathcal{C}$ becomes empty, the algorithm selects the $k$ pairs with the smallest distances in $\mathcal{R}$ and returns their vertices as $\mathcal{R}_k$ (Lines~24--25).

\section{Guided Search in UniGraph}
\label{sec:guided-uhg-search}

The guided search framework in UniGraph uses query-dependent entry points to initialize traversal of the unified graph. It selects between sparse-guided and dense-guided paths according to the query weight $\alpha$. We first describe how the two paths obtain their entry points, and then present the $\alpha$-adaptive routing policy and analyze the benefits of guided initialization.

\subsection{Guided Entry-Point Search}
\label{sec:guided-entry-search}

Guided UniGraph search uses a guide index to obtain query-dependent entry points, as illustrated by the concrete example in Figure~\ref{fig:uhg-search-example}. We use either a sparse index or a dense index as the guide index. The sparse- and dense-guided paths retrieve results under $D_s$ and $D_d$, respectively. The retrieved results serve as the entry-point set $E_q$ for initializing the traversal in Algorithm~\ref{alg:uhg-search}; thereafter, UniGraph ranks and explores vertices using the query-specific hybrid distance $D_h$. The framework is independent of the specific guide index implementation. Any dense or sparse index that returns results ranked by the corresponding component distance can serve as the guide index.

\subsection{$\alpha$-Adaptive Execution}

The guided search framework selects an execution path according to the query-level hybrid weight $\alpha$: sparse- and dense-dominant queries are routed to the sparse- and dense-guided paths, respectively. After routing, the corresponding sparse or dense guide index is queried, and its results form the entry-point set $E_q$ for UniGraph traversal. The two paths use the same UniGraph search procedure, differing only in the selected guide index and the resulting entry points.

We express this query-time selection as
\[
\operatorname{Path}(q,\alpha)=
\begin{cases}
\text{sparse-guided path}, & \alpha \leq \alpha_t,\\
\text{dense-guided path}, & \alpha > \alpha_t,
\end{cases}
\]
where $\alpha_t$ is the routing threshold between sparse-guided and dense-guided execution.

In our experiments, the dense and sparse distances have comparable scales. We therefore set $\alpha_t=0.5$, the balance point at which the two distances receive equal weights in Eq.~\eqref{eq:hybrid-distance}.

\subsection{Analysis}

Guided entry points improve search by reducing the number of hybrid graph traversal steps needed to reach the result region. During UniGraph traversal, visiting a candidate may require evaluating the hybrid distance, which combines both dense and sparse distances. In contrast, the sparse and dense guide indexes perform single-route coarse search using only one component distance. Since dense and sparse vectors are correlated, the top-$k$ results under either component distance are often close in UniGraph to highly ranked hybrid results. Guided UniGraph search can therefore be viewed as a coarse-to-fine process: it first uses $D_s$ or $D_d$ to approach the result region and then uses the query-specific hybrid distance $D_h$ to refine the search over UniGraph, reducing hybrid distance evaluations and improving search efficiency compared with unguided UniGraph traversal initialized from a random graph entry point.

The $\alpha$-adaptive policy further matches the entry-point strategy to the query's dense-sparse weighting. For smaller $\alpha$ values, results returned by the sparse guide index are more likely to include or lie near highly ranked hybrid results. For larger $\alpha$ values, results returned by the dense guide index provide analogous guidance for dense-dominant queries. Taken together, guided initialization and $\alpha$-adaptive routing provide two benefits: they reduce the hybrid distance evaluations needed to reach the result region and improve recall by selecting entry points that match the dense-sparse preference specified by the query.

\begin{table}[!t]
\centering
\caption{Dataset statistics. Avg. NNZ reports, in base/query order, the average number of nonzero entries per sparse vector.}
\label{tab:dataset-statistics}
\resizebox{\columnwidth}{!}{%
\begin{tabular}{lcccc}
\toprule
Dataset & Base & Queries & Dense Dim. & Avg. NNZ \\
\midrule
NQ & 2.68M & 1,000 & 1024 & 66.0/10.7 \\
HotpotQA & 5.23M & 1,000 & 1024 & 44.2/17.5 \\
MS MARCO & 8.84M & 1,000 & 1024 & 47.7/7.2 \\
FEVER & 5.42M & 1,000 & 1024 & 66.2/11.5 \\
DBpedia-Entity & 4.64M & 467 & 1024 & 47.8/7.3 \\
\bottomrule
\end{tabular}
}\vspace{-3ex}
\end{table}

\section{Experiments}
\label{sec:experimental}

\subsection{Experimental Setup}

We evaluate UniGraph in terms of retrieval effectiveness and efficiency under mixed and fixed query weights, as well as index construction time and storage cost. We further conduct ablation studies on guided entry points, hybrid distance pruning, and multi-weight graph construction.

\textit{Datasets, query weight settings, and metrics.}
We use five public retrieval datasets from the BEIR benchmark~\cite{thakur2021beir}: NQ~\cite{kwiatkowski2019natural}, HotpotQA~\cite{yang2018hotpotqa}, MS MARCO~\cite{bajaj2016msmarco}, FEVER~\cite{thorne2018fever}, and DBpedia-Entity~\cite{hasibi2017dbpedia}.
NQ covers open-domain question answering, HotpotQA focuses on multi-hop question answering, MS MARCO evaluates web passage retrieval, FEVER evaluates evidence retrieval for fact verification, and DBpedia-Entity evaluates entity retrieval over DBpedia articles.
Each document and query is encoded by BGE-M3~\cite{chen2024bgem3} into a 1024-dimensional dense vector and a weighted sparse vector.
The dense embeddings produced by BGE-M3 are $\ell_2$-normalized.
Table~\ref{tab:dataset-statistics} summarizes basic statistics for the encoded datasets.
For NQ, HotpotQA, MS MARCO, and FEVER, each query set contains 1,000 queries; for DBpedia-Entity, it contains all 467 available queries.
We evaluate both mixed-$\alpha$ and fixed-$\alpha$ query weight settings.
For each evaluated query--weight combination, we compute the exact hybrid top-$k$ result under Eq.~\eqref{eq:hybrid-distance} as ground truth.

We report Recall@50, Recall@100, and Recall@200 in the mixed-$\alpha$ setting, and Recall@100 in the fixed-$\alpha$ setting.
We measure end-to-end query throughput over the complete query set and report it in queries per second (QPS).

\begin{figure*}[!t]\vspace{0ex}
	\centering
	\includegraphics[width=\textwidth,trim={0pt 100pt 0pt 0pt},clip]{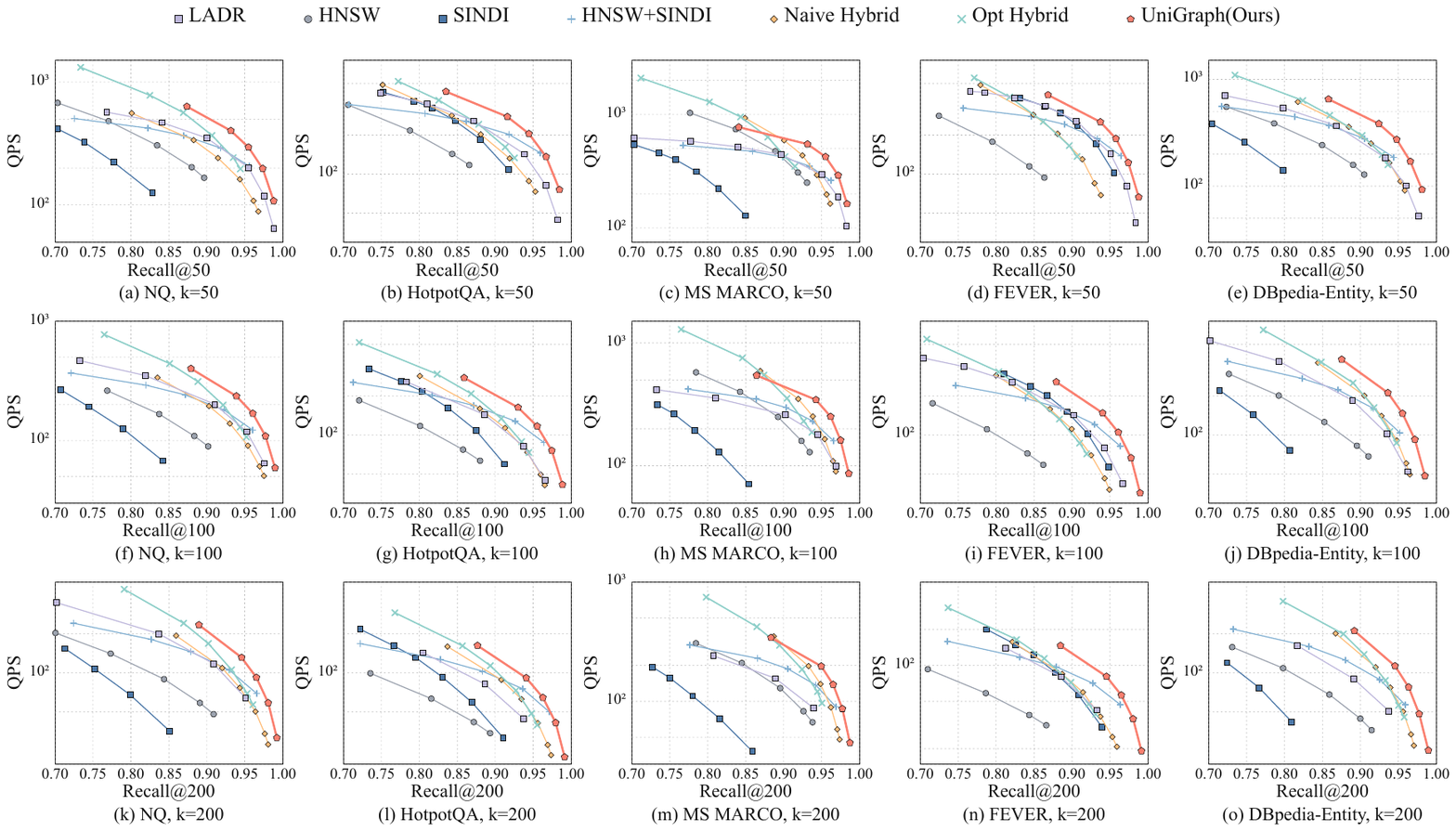}\vspace{-2ex}
	\caption{Recall@50-QPS (top), Recall@100-QPS (middle), and Recall@200-QPS (bottom) trade-offs in the mixed-$\alpha$ setting. The query weights are uniformly distributed over $[0.3,0.7]$. Curves closer to the upper-right corner indicate better performance.}\vspace{-1ex}
	\label{fig:mixed-alpha-results}
\end{figure*}

\textit{Baselines.}
We compare UniGraph with six baselines.
Across all evaluated methods, we use HNSW~\cite{malkov2020hnsw} and SINDI~\cite{li2026sindi} as the dense and sparse indexes, respectively.
HNSW is a widely used graph-based ANNS index that organizes dense vectors in a hierarchical proximity graph, while SINDI is an efficient inverted-index-based sparse retrieval method without redundant computation.
Below, we describe each baseline.
We reimplement the Naive Hybrid and Opt Hybrid variants introduced in~\cite{zhang2024hybridann}.
These two baselines represent fixed-weight hybrid graph approaches, whereas HNSW+SINDI represents the two-route retrieval approach.

\begin{table}[!t]
\centering
\caption{Fixed parameter settings for index construction and search.}
\label{tab:parameter-settings}
\small
\renewcommand{\arraystretch}{1.08}
\setlength{\tabcolsep}{5pt}
\begin{tabularx}{\columnwidth}{@{}l>{\raggedright\arraybackslash}X@{}}
\toprule
Method & Parameter Settings \\
\midrule
HNSW
& \textit{Construction:} $M=64$; construction beam size $ef_c=200$\newline
\textit{Search:} Candidate count $=ef_s$ \\
\midrule
SINDI
& \textit{Construction:} Document pruning ratio $=0$\newline
\textit{Search:} Query pruning ratio $=0.5$ \\
\midrule
HNSW+SINDI
& \textit{Construction:} HNSW and SINDI settings as specified above\newline
\textit{Search:} Dense candidate count $=$ sparse candidate count; SINDI query pruning ratio $=0.5$ \\
\midrule
Naive Hybrid
& \textit{Construction:} $\alpha=0.5$; $M=64$; $ef_c=200$ \\
\midrule
Opt Hybrid
& \textit{Construction:} $\alpha=0.5$; $M=64$; $ef_c=200$; hybrid neighbor expansion width $=200$\newline
\textit{Search:} dense-stage stopping threshold $\tau_d=0.8$; hybrid-stage stopping threshold $\tau_h=1.0$; no sparse vector pruning \\
\midrule
LADR
& \textit{Construction:} HNSW and SINDI settings as specified above\newline
\textit{Search:} Lexical seed count $=200$; SINDI query pruning ratio $=0.5$ \\
\midrule
UniGraph
& \textit{Construction:} $\mathcal{A}=\{0.0,0.1,\ldots,1.0\}$; dense/sparse candidate counts $=100/100$; HNSW $ef_s=200$; SINDI query pruning ratio $=0.9$; $k_g=32$; $M=64$\newline
\textit{Search:} $\gamma=0.3$; guided-entry count $=$ guide-HNSW $ef_s$ $=$ UniGraph $ef_s$; SINDI query pruning ratio $=0.5$ \\
\bottomrule
\end{tabularx}\vspace{-2ex}
\end{table}

\begin{itemize}
    \item \textit{LADR:}
    LADR~\cite{kulkarni2023ladr} uses sparse retrieval results as entry points and, with an adaptive expansion width ($c$), repeatedly expands the neighbors of the current top-$c$ candidates ranked by their dense scores until convergence.
    We rerank all candidates using the query-specific hybrid distance.

    \item \textit{HNSW (dense-only) and SINDI (sparse-only):}
    HNSW and SINDI retrieve candidates according to the dense and sparse distances, respectively.
    Each baseline then computes the missing distance component for its retrieved candidates and reranks them using the query-specific hybrid distance.

    \item \textit{HNSW+SINDI (two-route):}
    This baseline represents the two-route retrieval approach.
    It independently retrieves candidate sets from HNSW and SINDI for dense and sparse vectors respectively, and reranks the resulting candidates using the query-specific hybrid distance.

    \item \textit{Naive Hybrid~\cite{zhang2024hybridann}:}
    The Naive Hybrid variant constructs and searches an HNSW index directly using the hybrid distance with a fixed weight of $\alpha=0.5$.

    \item \textit{Opt Hybrid~\cite{zhang2024hybridann}:}
    The Opt Hybrid variant starts from a dense HNSW graph, augments it with neighbors selected by the fixed-weight hybrid distance, and performs dense traversal followed by hybrid traversal.
    Although the original method includes sparse vector pruning, we disable this optimization because our preliminary experiments found that it yielded only marginal end-to-end throughput improvements under the evaluated workloads and did not materially change the Recall-QPS trade-off.
\end{itemize}

\textit{UniGraph} selects SINDI or HNSW according to the query weight to obtain entry points, then traverses the unified graph and ranks candidates using the query-specific hybrid distance.
UniGraph is compatible with other dense and sparse indexes.

\begin{figure*}[!t]\vspace{0ex}
	\centering
	\includegraphics[width=\textwidth,trim={0pt 100pt 0pt 0pt},clip]{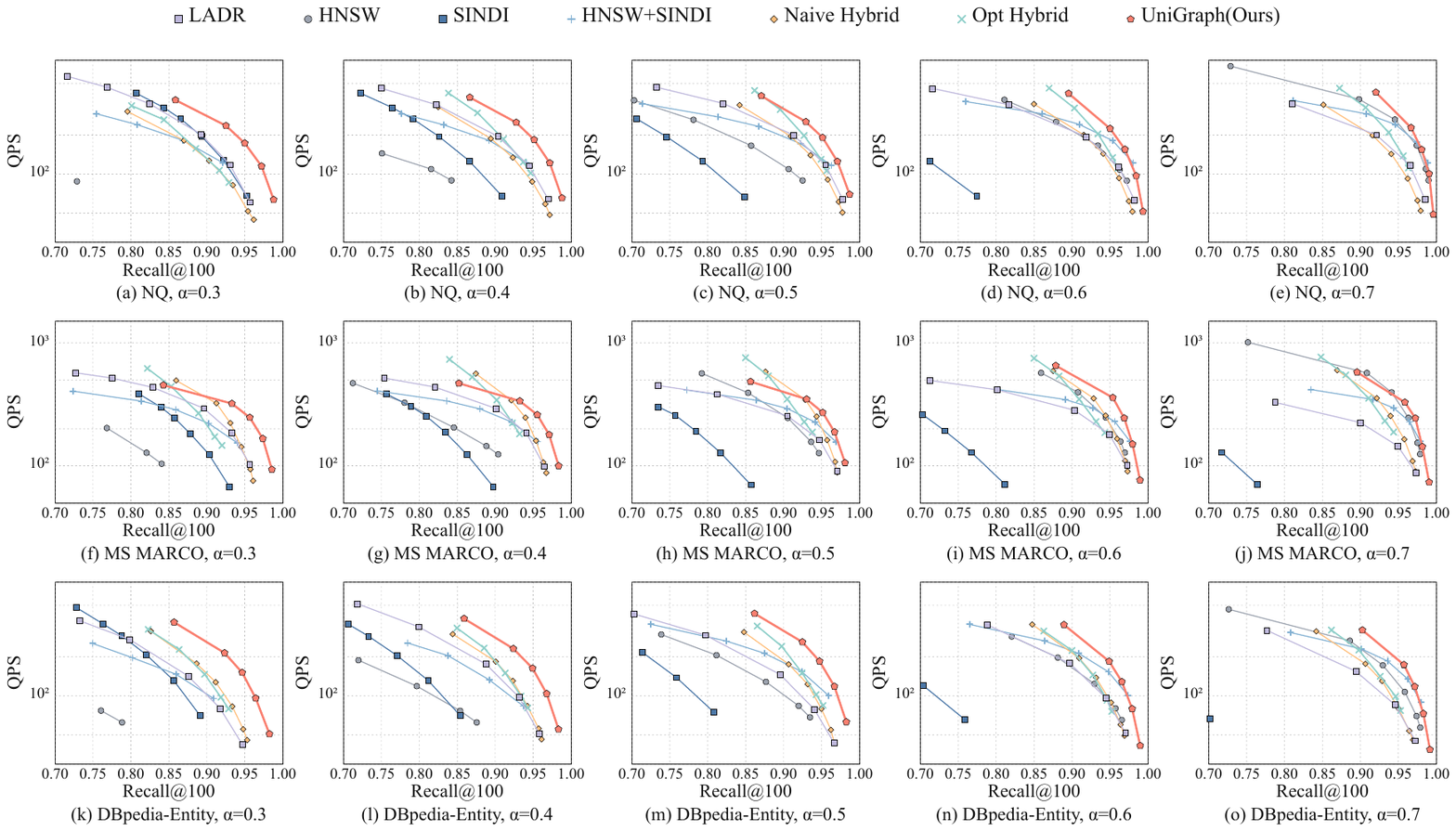}\vspace{-2ex}
	\caption{Fixed-$\alpha$ Recall@100-QPS trade-offs on NQ, MS MARCO, and DBpedia-Entity. The upper-right region is better.}\vspace{-1ex}
	\label{fig:fixed-alpha-results}
\end{figure*}

\textit{Parameter settings.}
UniGraph uses HNSW and SINDI as dense and sparse candidate generators, respectively, during graph construction.
The Naive Hybrid and Opt Hybrid baselines are constructed at the fixed weight $\alpha=0.5$, while LADR uses a proximity graph extracted from the dense HNSW index and obtains its lexical seeds from SINDI.

For the Recall-QPS curves, the graph search width ranges approximately from $k$ to $10k$ for HNSW, the two fixed-weight hybrid baselines, and UniGraph.
The SINDI candidate count ranges from $10k$ to $100k$. For the two-route baseline, the dense and sparse candidate counts are set equal, as determined by grid search, and the count for each route ranges from $k$ to $5k$.
For LADR, the adaptive expansion width ranges from 1 to 200.
Table~\ref{tab:parameter-settings} summarizes the settings.

\textit{Experimental environment.}
All methods are implemented in C++17 and compiled with GCC/G++ 15.2.0 using CMake 4.2.3. All experiments are conducted on a server running Linux 5.10.134 and equipped with one AMD EPYC 9T24 processor with 96 physical cores and 1.1\,TiB of memory. Index construction uses up to 32 physical cores. At search time, all indexes and vector data reside entirely in memory, and all search experiments use a single thread.

\subsection{Overall Performance}

\stitle{1) Mixed-$\alpha$ Setting.}
The mixed-$\alpha$ setting tests whether an index remains effective when the relative weighting of dense and sparse signals varies from query to query. For each dataset, the query weights are uniformly distributed over $[0.3,0.7]$. We select this interval, because at more extreme weights such as $\alpha=0.1$ or $0.9$, the hybrid distance is determined primarily by only one modality.

Figure~\ref{fig:mixed-alpha-results} compares the Recall-QPS trade-offs of all methods when $\alpha$ varies across queries.
At Recall@100 $\geq0.95$, UniGraph achieves 27.6\%$\sim$52.3\% higher QPS than the fastest baseline across the five datasets.
On NQ, UniGraph sustains about $1.7\times$ the QPS of two-route at Recall@100 $\approx 0.88$.
A more pronounced advantage appears on FEVER, where UniGraph delivers about $3.0\times$ the QPS of Naive Hybrid at Recall@100 $\approx 0.94$.
UniGraph also retains a clear advantage over LADR on DBpedia-Entity, providing an approximately $2.3\times$ QPS speedup at Recall@100 $\approx 0.95$.

\stitle{2) Fixed-$\alpha$ Setting.}
The fixed-$\alpha$ setting evaluates performance when all queries share the same weight. We consider five fixed weights: $\alpha\in\{0.3,0.4,0.5,0.6,0.7\}$.

Figure~\ref{fig:fixed-alpha-results} reports the Recall@100-QPS trade-offs at each fixed $\alpha$ value.
Among the two single-modality baselines, the sparse-only baseline achieves higher QPS at matched recall for smaller $\alpha$, whereas the dense-only baseline achieves higher QPS for larger $\alpha$.
At Recall@100 $\geq0.95$, UniGraph achieves $1.18\times$$\sim$$2.85\times$ the QPS of the fastest baseline, with a geometric-mean speedup of $1.66\times$ over all three datasets and five query weights.

On NQ with $\alpha=0.3$, UniGraph delivers about $2.5\times$ the QPS of the sparse-only baseline at Recall@100 $\approx 0.95$.
Compared with the fixed-weight hybrid graph baselines, UniGraph achieves $1.44\times$$\sim$$3.34\times$ the QPS of Naive Hybrid and $1.40\times$$\sim$$2.26\times$ that of Opt Hybrid in their comparable configurations. The advantage persists at $\alpha=0.5$, even though this query weight matches the construction weight of both Hybrid indexes.
On MS MARCO with $\alpha=0.6$, UniGraph also provides about $2.1\times$ the QPS of LADR at Recall@100 $\approx 0.95$, consistent with the benefit of traversing the graph under the query-specific hybrid distance rather than dense scores alone.
At $\alpha=0.5$ on DBpedia-Entity, UniGraph maintains about $1.6\times$ the QPS of two-route at comparable recall.

\begin{figure}[t]
\centering
\includegraphics[width=\columnwidth,trim={0pt 250pt 0pt 0pt},clip]{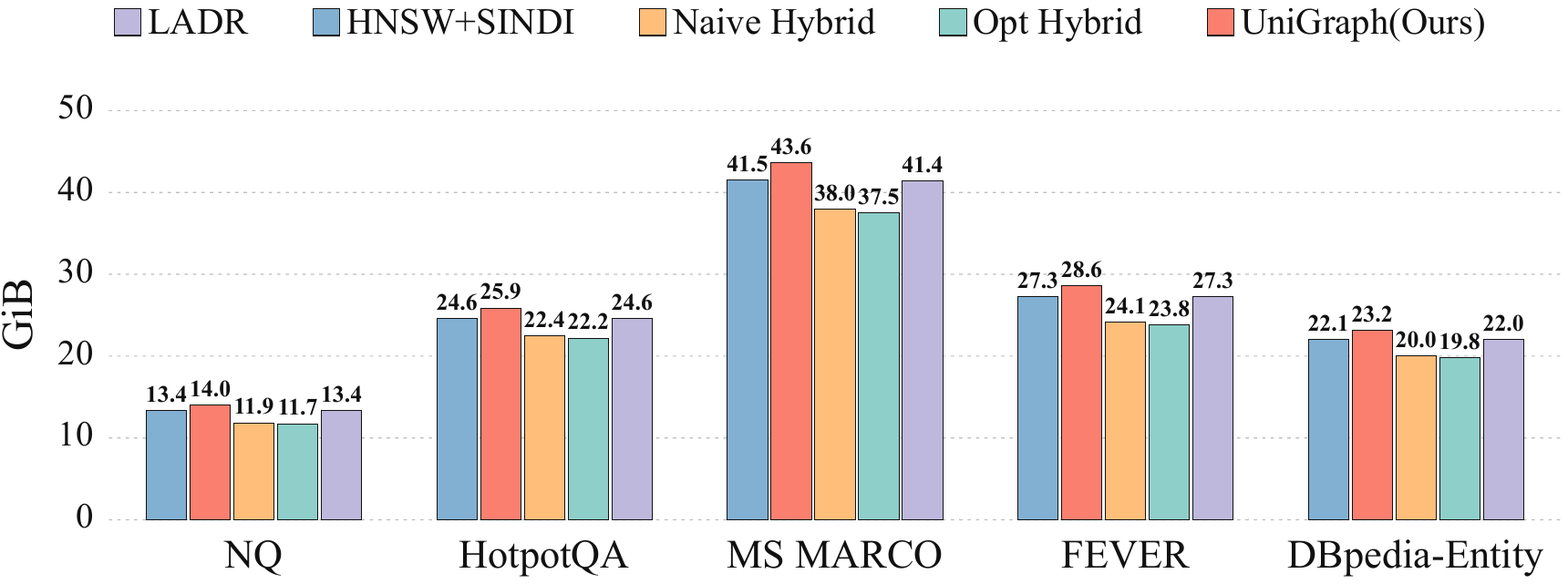}\vspace{-2ex}
\caption{Index size of HNSW+SINDI, UniGraph, Naive Hybrid, Opt Hybrid, and LADR on all five datasets; lower is better.}\vspace{-1ex}
\label{fig:index-size}
\end{figure}

\begin{figure*}[t]
\centering
\includegraphics[width=\textwidth,trim={15pt 370pt 20pt 0pt},clip]{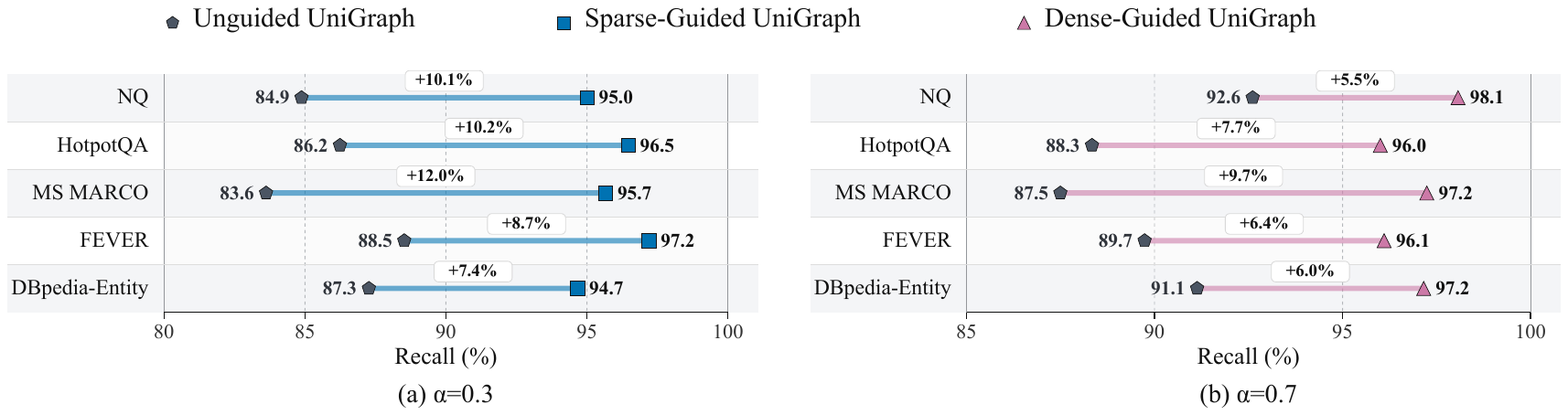}\vspace{-1ex}
\caption{Recall@100 of unguided and guided UniGraph search at $ef_s=300$. Panel (a) compares unguided and sparse-guided UniGraph search at $\alpha=0.3$, while panel (b) compares unguided and dense-guided UniGraph search at $\alpha=0.7$.}
\label{fig:entry-ablation}\vspace{-1ex}
\end{figure*}

\stitle{3) Index Construction and Storage.}
Table~\ref{tab:construction-time} compares the construction times of two-route, UniGraph, Naive Hybrid, Opt Hybrid, and LADR, while Figure~\ref{fig:index-size} compares their index sizes across all five datasets. Building the unified graph introduces additional offline construction cost relative to two-route. Nevertheless, UniGraph requires 6.6\%$\sim$21.9\% less construction time than Opt Hybrid and incurs only 4.5\%$\sim$5.3\% additional storage over two-route.

The reported construction time includes all index components required by each method. For UniGraph, it includes constructing HNSW, SINDI, and the unified graph.

\begin{table}[t]
\centering
\caption{Index construction time in seconds; lower is better.}
\label{tab:construction-time}
\scriptsize
\renewcommand{\arraystretch}{1.08}
\setlength{\tabcolsep}{2.5pt}
\begin{tabular*}{\columnwidth}{@{\extracolsep{\fill}}lrrrrr@{}}
\toprule
Dataset & \multicolumn{1}{c}{\shortstack{HNSW+\\SINDI}} &
\multicolumn{1}{c}{UniGraph} &
\multicolumn{1}{c}{\shortstack{Naive\\Hybrid}} &
\multicolumn{1}{c}{\shortstack{Opt\\Hybrid}} &
\multicolumn{1}{c}{LADR} \\
\midrule
NQ & \textbf{438} & 918 & 910 & 1,175 & 442 \\
HotpotQA & \textbf{1,181} & 2,497 & 1,592 & 3,069 & 1,190 \\
MS MARCO & \textbf{1,418} & 2,946 & 2,265 & 3,155 & 1,428 \\
FEVER & \textbf{1,300} & 2,844 & 2,060 & 3,609 & 1,309 \\
DBpedia-Entity & \textbf{1,052} & 2,215 & 1,450 & 2,752 & 1,060 \\
\bottomrule
\end{tabular*}
\end{table}

The construction cost of HNSW+SINDI is dominated by its HNSW component. In our measurements, SINDI takes only tens of seconds to build, so the construction time of HNSW+SINDI remains close to that of HNSW alone. Naive Hybrid incurs additional cost by evaluating the hybrid distance directly during graph construction. Opt Hybrid further augments a dense HNSW graph with neighbors selected by the hybrid distance, while UniGraph constructs neighborhoods at multiple weights and merges them into a unified graph. These additional hybrid distance evaluations and graph processing steps account for the higher construction costs of the graph-based hybrid methods.

Compared with Naive Hybrid, Opt Hybrid, and LADR, UniGraph uses 14.7\%$\sim$18.7\%, 16.3\%$\sim$20.2\%, and 4.5\%$\sim$5.5\% more index space, respectively. The additional space is the cost of maintaining both the multi-weight unified graph and the two guide structures. By sharing the underlying vectors and labels among these structures, UniGraph limits its storage overhead relative to the two-route baseline to about 5\% while supporting efficient retrieval under query-dependent hybrid weights without graph reconstruction.

\begin{figure}[t]
\centering
\includegraphics[width=\columnwidth,trim={40pt 180pt 45pt 20pt},clip]{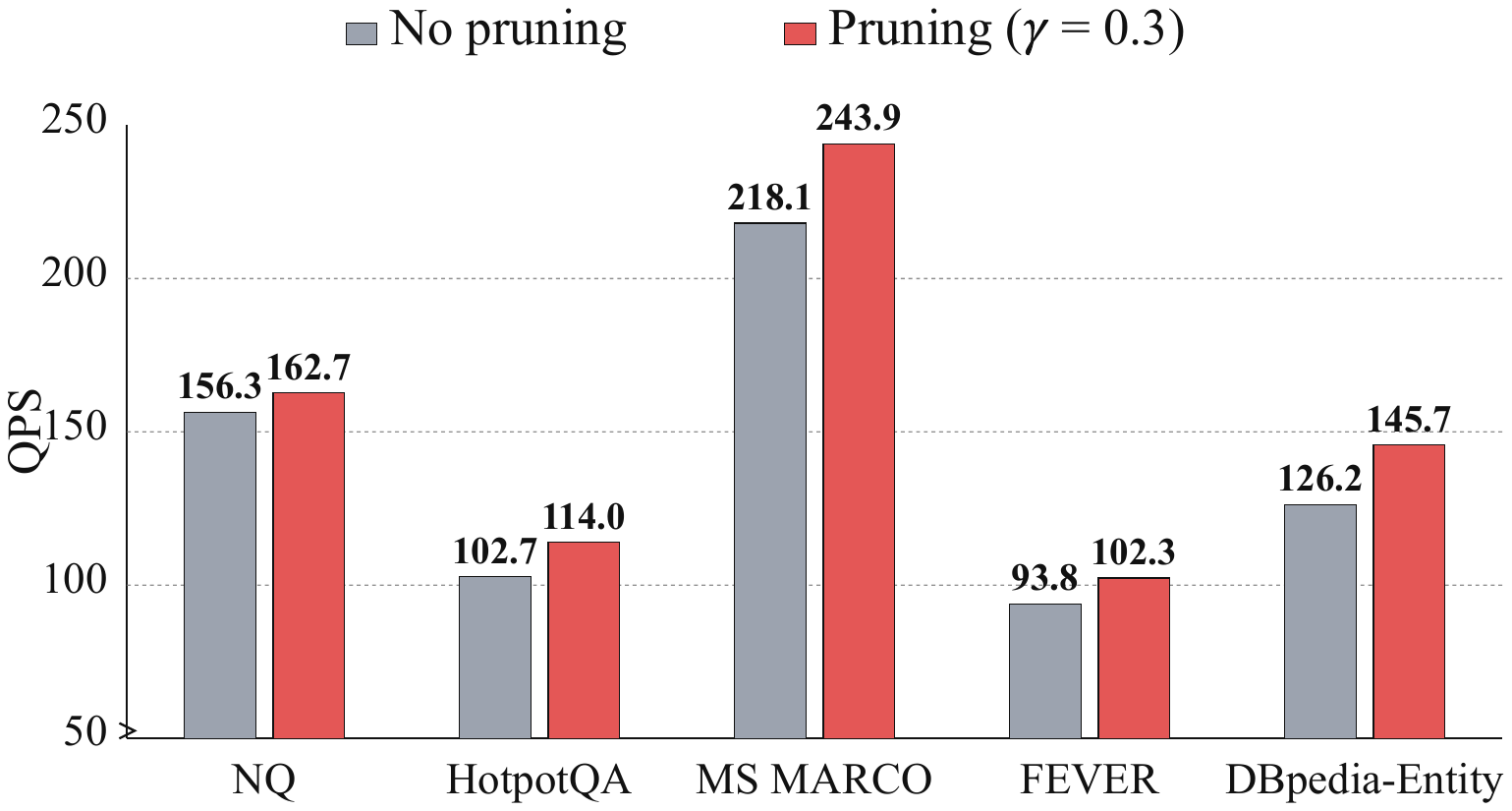}\vspace{-1ex}
\caption{QPS with and without dense-first hybrid distance pruning under the mixed-$\alpha$ setting at $ef_s=300$.}\vspace{-1ex}
\label{fig:pruning-ablation}
\end{figure}

\subsection{Ablation Study}

We evaluate how guided entry points affect retrieval effectiveness, how hybrid distance pruning affects search efficiency, and whether multi-weight graph construction improves the graph-only Recall-QPS trade-off.

\stitle{1) Entry-Point Strategy.}
We compare unguided UniGraph search initialized from a random graph entry point with sparse-guided UniGraph search at $\alpha=0.3$ and dense-guided UniGraph search at $\alpha=0.7$. All variants use the same hybrid graph, pruning scale $\gamma=0.3$, and search beam size $ef_s=300$.

Figure~\ref{fig:entry-ablation} shows that guided entry points substantially improve recall in all comparisons. Unguided UniGraph search obtains 83.6\%$\sim$92.6\% Recall@100, whereas the sparse- and dense-guided configurations reach 94.7\%$\sim$98.1\%. At $\alpha=0.3$, sparse-guided UniGraph search improves recall by 7.4$\sim$12.0 percentage points, averaging 9.7 points across the datasets. At $\alpha=0.7$, dense-guided UniGraph search improves recall by 5.5$\sim$9.7 percentage points, averaging 7.1 points. At both tested weights, using the corresponding sparse or dense guide index to select entry points substantially improves recall over random initialization.

\stitle{2) Hybrid Distance Pruning.}
For the hybrid distance pruning ablation, we use the mixed-$\alpha$ setting to compare UniGraph search with dense-first hybrid distance pruning ($\gamma=0.3$) and UniGraph search without pruning. To isolate the effect of pruning, both configurations use the same prebuilt UniGraph and select between sparse-guided and dense-guided search using the same $\alpha$-adaptive routing policy. We set $ef_s=300$. The configuration without pruning disables only the pruning test in Eq.~\eqref{eq:pruning-bound}; candidate insertion and traversal termination still use the current result set threshold $\tau$, as in standard beam search.

As Figure~\ref{fig:pruning-ablation} shows, hybrid distance pruning improves QPS on all datasets, with an average improvement of 10.3\%. This throughput improvement incurs negligible recall loss: the maximum decrease in Recall@100 is only 0.055 percentage points. These results demonstrate that hybrid distance pruning effectively accelerates UniGraph search with almost no impact on result quality.

\stitle{3) Multi-Weight Graph Construction.}
To examine the effect of multi-weight graph construction, we compare UniGraph with a fixed-weight graph, which selects nearest neighbors only at $\alpha=0.5$ for graph construction with $M=64$. Query weights are uniformly distributed over $[0,1]$. To exclude the effects of the query-time optimizations studied above, both methods use graph-only search initialized from the same randomly selected entry point for each query, with dense/sparse-guided entry points and hybrid distance pruning disabled.

\begin{table}[t]
\centering
\caption{Graph-only QPS of the fixed-weight graph and UniGraph at matched Recall@100; higher is better.}
\label{tab:multi-weight-construction-ablation}
\small
\renewcommand{\arraystretch}{1.08}
\setlength{\tabcolsep}{3.5pt}
\begin{tabular*}{\columnwidth}{@{\extracolsep{\fill}}lrrrr@{}}
\toprule
Dataset & Recall@100 & \shortstack{Fixed-weight\\graph} & UniGraph & Gain \\
\midrule
MS MARCO & 0.90 & 150.96 & \textbf{189.39} & +25.46\% \\
MS MARCO & 0.95 & 53.88 & \textbf{68.28} & +26.71\% \\
FEVER & 0.90 & 92.73 & \textbf{104.85} & +13.07\% \\
FEVER & 0.95 & 35.97 & \textbf{41.77} & +16.13\% \\
DBpedia-Entity & 0.90 & 129.57 & \textbf{139.87} & +7.95\% \\
DBpedia-Entity & 0.95 & 48.39 & \textbf{50.55} & +4.48\% \\
\bottomrule
\end{tabular*}\vspace{-3ex}
\end{table}

As shown in Table~\ref{tab:multi-weight-construction-ablation}, UniGraph achieves higher QPS than the fixed-weight graph on all three datasets at both target recall levels. At Recall@100 $=0.90$, its QPS improvements range from 7.95\% to 25.46\%, with a mean improvement of 15.49\%. At Recall@100 $=0.95$, the improvements range from 4.48\% to 26.71\%, averaging 15.77\%. These results show that, under the graph-only setting, constructing one graph from neighborhoods sampled at multiple weights provides a better high-recall throughput trade-off than constructing the graph at a single fixed weight.

\section{Related Work}
\label{sec:related}

\stitle{Fusion-based dense-sparse retrieval.}
Fusion-based systems typically retrieve dense and sparse candidates separately and then combine their truncated result lists. Weighted score fusion~\cite{bruch2024fusion} preserves score magnitudes but requires calibration, whereas reciprocal rank fusion (RRF)~\cite{cormack2009rrf,bruch2024fusion} avoids calibration but discards score magnitudes and is sensitive to its rank constant. They both can lose candidates before fusion.

\stitle{Dense and sparse retrieval.}
Dense vector ANNS methods include locality-sensitive hashing~\cite{indyk1998ann,gionis1999lsh}, quantization and inverted file methods (e.g, product quantization~\cite{jegou2011pq} and IVF-PQ~\cite{johnson2019faiss}), and graph-based indexes (e.g., HNSW~\cite{malkov2020hnsw}, NSG~\cite{fu2019nsg}, and DiskANN~\cite{subramanya2019diskann}). Empirical evaluations show that graph-based methods offer competitive recall-efficiency trade-offs and often perform particularly well in high-recall regimes~\cite{aumuller2020annbenchmarks,wang2021graphsurvey}.

Efficient sparse retrieval is commonly implemented with inverted indexes. Representative approaches accelerate retrieval through pruning or compact index structures. Classical block-max pruning and BMP skip low-potential document blocks~\cite{ding2011blockmax,mallia2024blockmax}; Seismic groups inverted lists into geometrically cohesive blocks~\cite{bruch2024efficient}; Sinnamon uses compact sketches~\cite{bruch2023streaming}; and SINDI uses optimized inverted lists for approximate sparse maximum inner-product search~\cite{li2026sindi}. 
%In our two-route setting, HNSW and SINDI represent the dense and sparse routes, respectively. 
%Both are effective single-modality retrievers, but neither alone is well suited to hybrid retrieval.

\stitle{Unified dense-sparse indexing and dynamic-weight hybrid search.}
A graph-based ANNS method for dense-sparse hybrid vectors~\cite{zhang2024hybridann} aligns the dense and sparse distance distributions and refines an HNSW-style topology under a predetermined hybrid weight. 
Its fixed-weight topology may not match queries with different weights.

Existing query-dependent hybrid methods face limitations in dense-sparse retrieval. IVF-MIPS~\cite{bruch2024bridging} reuses IVF partitions for query-side reweighting, but lacks cross-weight neighborhood optimization. LADR~\cite{kulkarni2023ladr} seeds dense graph exploration with lexical results, but is tailored for dense retrieval. 
While GPU-based Allan-Poe~\cite{li2025allanpoe} supports dynamic multi-path weighting without index reconstruction, UniGraph builds a CPU-based bounded-degree graph by merging multi-weight sampled neighborhoods.

Other multi-metric approaches and DEG~\cite{yin2025deg} support query-specific weights via Pareto frontier neighborhoods or multi-graph traversals~\cite{franzke2016multimetric,zhu2022desire,hu2020wlsh,zhu2024hjg,uemura2024multimetric}. However, DEG is restricted to fixed-length Euclidean embeddings; its formulation and analysis do not cover high-dimensional inner-product sparse vectors, making it inapplicable to our dense-sparse hybrid setting without fundamental redesign.

\section{Conclusion}
\label{sec:conclusion}

This paper studies efficient dense-sparse hybrid retrieval with query-specific hybrid weights. We investigate UniGraph, a unified hybrid graph that supports different query weights without index reconstruction. UniGraph uses an $\alpha$-cover construction to merge neighborhoods sampled at multiple weights. We further study $\alpha$-adaptive graph search with sparse-guided and dense-guided execution paths. We also study dense-first hybrid distance pruning for reducing exact sparse distance computations. Experiments show that, at Recall@100 $\geq 0.95$, UniGraph achieves 27.6\%$\sim$52.3\% higher QPS than the fastest baseline on each dataset under mixed weights and $1.18\times$$\sim$$2.85\times$ QPS of the fastest baseline across the fixed-weight settings.

%\section*{Acknowledgment}
%The authors thank the reviewers and collaborators for their constructive feedback.

\newpage
\section*{Acknowledgment}
The authors disclose the use of OpenAI's ChatGPT to assist with drafting
the experimental code.  The experimental design,
testing, and correctness validation of all results were performed by the
authors.  ChatGPT was additionally used for editing and grammar polishing
of parts of the manuscript.  All AI-assisted content was reviewed and
verified by the authors, who take full responsibility for the
correctness and originality of this paper.

\bibliographystyle{IEEEtran}
\bibliography{references}

\end{document}